\documentclass[preprint]{aastex}

\begin{document}
\title{Optical and Infrared Photometry of the Micro-quasar GRO J1655-40 in 
Quiescence}
\author{Jenny Greene, Charles D. Bailyn}
\affil{Astronomy Department, Yale University, PO Box 208101,
New Haven CT, 06520-8101, USA}
\email{jenny.greene@aya.yale.edu; bailyn@astro.yale.edu}

\author{Jerome A. Orosz}
\affil{Astronomical Institute, Utrecht University, PO Box 80000, 
3508 TA Utrecht, The Netherlands}
\email{J.A.Orosz@astro.uu.nl}

\begin{abstract}

We present $BVIJK$ photometry of the black--hole candidate GRO J1655--40 
in full quiescence.  We report a refined orbital period of 
$2.62191 \pm 0.00020$ days.  
The light curves are dominated by ellipsoidal variations 
from the secondary star.  
We model the light curves with an upgraded code which 
includes a more accurate treatment of limb darkening.
Previous models 
containing a large cool disk are ruled out, and indeed our 
data can be fit with a 
pure ellipsoidal light curve without any disk contribution.
In general agreement with previous results, 
we derive a confidence region of the 
correlated quantities inclination and mass ratio, centered on an inclination
of $70.2 \pm 1.9 \arcdeg$ and mass ratio 
of $2.6 \pm 0.3$, resulting in a primary mass 
$M_1=6.3 \pm0.5\,M_{\sun}$ (all $95\%$ confidence).
The complex limits and errors on these values, and on the possible disk
contribution to the light curve, are discussed.

\end{abstract}


\date {}

\section{Introduction}

The Soft X-ray Transients (SXT) are a sub-class of 
accreting low-mass binary systems consisting of a neutron star or black hole
primary, accretion disk, and Roche lobe filling secondary star.  
Instabilities in mass accretion cause these
sources to undergo episodic outbursts.  The X-ray emission during outburst
indicates the existence of an accreting compact object.  During 
outburst, the accretion flow contributes nearly 
all the optical flux.  
In quiescence, on the other hand, the X-ray luminosity is very low, and
the optical light is dominated by the contribution of the secondary star.  
At this time the orbital characteristics of the SXT's may be studied in
some detail \citep{vPM95}.  In particular, the mass function can be measured, 
which provides a lower limit to the mass of the compact object.  
Further information about the mass ratio (from line broadening) and 
inclination (from ellipsoidal variability) can be used to determine the 
mass of the compact object \citep[and references therein]{BJCO98}
which in many cases is above the 
$3\,M_{\sun}$ upper limit of the mass of a neutron star \citep{CH76}. 

While this procedure has been carried out successfully for several systems 
(see references in Bailyn et al. 1998),
the modeling of ellipsoidal variations can be  
complicated by various other contributions to the observed light curves, 
in particular contributions from the accreting material.  The 
accretion flow typically contributes both a constant flux offset and 
random flickering.
The constant flux contribution of the disk decreases the observed 
amplitude of the 
ellipsoidal variations, and is thus degenerate with the inclination to 
first order (there are potentially
observable shape differences).  In systems
for which the disk contribution is significant, not only is the amplitude
of the variations decreased, but the aperiodic flickering may result in changes
in the light curve from measurement to measurement, complicating the process
of accurately modeling the light curve. 
One proposed method to minimize the disk contribution
is the use of infrared light curves.  For example, \citet{SNC94} suggest 
that the disk contribution to the light curves of A0620--00 are
 minimized in the infrared.  They have directly measured the disk contribution
in the $K$--band spectroscopically, and find it to be less than 
$27\%$, smaller than in the optical \citep{SBC99}.

Models used to fit the ellipsoidal variations of SXT's in quiescence
have traditionally assumed accretion flow in the form of a geometrically thin, 
optically thick disk (e.g. Orosz and Bailyn 1997, hereafter OB97; 
van der Hooft et al. 1998).  
The standard thin $\alpha $-disk model corresponds to one
allowed solution to the hydrodynamic equations which describe 
differentially rotating gas flows \citep{SS73,FKR92}.  
The $\alpha$-disk model has
been successfully used to model the outburst cycle of cataclysmic variables
({\it e.g.} Cannizzo 1993).  
However, recent findings have called into question 
the application of standard thin
disk to model SXT systems.  In particular, Narayan and collaborators have 
suggested that the accretion flow in quiescence consists of an outer
disk and an optically thin advection dominated flow closer to the compact
object \citep{NMY96}.

The SXT GRO J1655--40 (Nova Sco 1994) was discovered on 
July 7 1994 with the Burst and Transient Source Experiment 
aboard the {\it Compton Gamma Ray Observatory} \citep{Zh94}. 
Because GRO J1655--40 has a relatively luminous F star secondary the 
quiescent light curves 
are dominated by star light to an unusually large extent.  As a result, an 
unusually precise
fit of the orbital parameters is possible.
OB97 report values of 
$P_{orb}=2.62157 \pm 0.00015$ days, 
$f(M)=3.24 \pm 0.09 \,M_{\sun}$,
${\it i}=69.50 \pm 0.08{\arcdeg}$, and 
$Q=2.99 \pm 0.08$ (where $Q$ is taken to be
$M_1/M_2$)  
resulting in a primary mass of 
$7.09 \pm 0.22 \,M_{\sun}$.  \citet{vdH98} derive a refined 
$P_{orb}=2.62168 \pm 0.00014$ 
days from photometric and spectroscopic quiescent data taken before the 
April 1996 outburst.  They use  
$f(M)=3.16 \pm 0.15$ \citep{BOMR95b}, 
and report a range of secondary mass
$1.60 \la M_2 \la 3.10 \,M_{\sun}$ and {\it i} of 
$63.7 \la {\it i} \la 70.7{\arcdeg}$, 
corresponding to a primary mass range of $6.29-7.60 \,M_{\sun}$.
  \citet{ShvdH99} obtained improved spectroscopy,
and report $f(M)=2.73 \pm 0.09 \,M_{\sun}$ and a range of $q=M_2/M_1$ of 
$0.337-0.436$.  
They derive a primary mass of $5.5-7.9 \,M_{\sun}$.
Using Keck spectroscopy,
\citet{Isr99} found a rotational velocity for the secondary star of
v$_{\rm rot}\sin i=93\pm 3$ km s$^{-1}$, which implies a mass ratio of
$Q$= $2.48 \pm 0.16$ (assuming synchronous rotation and 
using the Shahbaz et al. mass function).
No infrared data have been available to date.

We have obtained $BVIJK$ light curves in order to explore the nature 
of the quiescent accretion flow and to further constrain the 
orbital parameters of the system.  In \S 2 we
present the data and reduction techniques.  In \S 3 we discuss the values
and limits on the orbital parameters and disk contribution.  Our results 
are summarized in \S 4.

\section{Observations and Reductions}

Photometry of the source was obtained from 1999 July through October with 
ANDICAM \footnote{http://www.astronomy.ohio-state.edu/~depoy/research/instrumentation/andicam/andicam.html}
on the 1.0 meter Yale telescope at CTIO, operated by the 
YALO Consortium \citep{BDA99}.  ANDICAM is 
capable of observing at an optical and an infrared wavelength
simultaneously.  The source was observed 
in the Johnson $BVI$ filters with a Loral $2048 \times 2048$ CCD and the
$JK$ infrared filters with a Rockwell $1024 \times 1024$ HgCdTe HAWAII Array.  
The infrared channel of ANDICAM allows for a dither pattern of up to seven
positions.  An observing sequence consisted of a single 720 second 
exposure in $B$ simultaneously with seven dithered 100 second exposures in 
$J$, followed by a single 360 second exposure in $V$ simultaneously with 
seven dithered 45 second exposures in $K$, followed by a single 180
second exposure in $I$ simultaneously with 4 dithered 45 second exposures 
in $K$.
We typically performed
one observing sequence per night between 
1999 July 7 and 1999 October 30 (HJD 2,451,376--2,451,481).

The optical images were bias and flat--field corrected using standard 
IRAF routines.  Sky corrections for the infrared frames were constructed by
median-combining all frames from a given night of the same filter.  This sky
field was subtracted from each image in the set.  A flat--field was 
constructed 
by taking the difference between a dome flat taken with dome--lights on, and 
one taken with dome--lights off.  Finally, the images from each set were 
co--added.

The program DAOPHOT II \citep{S87} within IRAF was used to 
remove near neighbors 
of the program object and all comparison stars.  Aperture photometry was then 
performed on the cleaned images.  The magnitudes of at least 30 nearby stable
stars were combined to form a comparison sequence for the 
differential photometry.
Time-series were constructed in each filter (see Figure 1).  In order to 
estimate the accuracy of our photometry, we extracted $\sim 5$ stable
comparison stars which span the magnitude range of the source.  We measured the
scatter of each comparison star about its mean, and use the average rms
of all the comparison stars as our estimate for the scatter in the 
program object photometry.  We find the average rms of these 
comparison stars to be $0.017$ mag in the
$B$ and $V$ filters, 0.014 mag in the $I$, $0.023$ mag in $J$ and 
0.032 mag in $K$.

Our $BVIJK$ light curves are presented in Figure 1 phased on our current best 
period of $2.62191 \pm 0.00020$ days (1 $\sigma $ error).  
They are plotted with a stable comparison star
of similar magnitude in order to demonstrate the
scatter in the photometry.
The data have been phased following the convention of OB97
in which phase 0 corresponds to the closest approach of the secondary 
star to the observer.  The light curves are dominated by
ellipsoidal variations, demonstrating two minima per orbit.  We observe a
deeper minimum at phase 0.5, which is explained by increased gravity
darkening near the inner Lagrangian point.  The optical light curves 
are quantitatively indistinguishable from those reported by OB97 
and \citet{vdH98}.  This result 
confirms that the data obtained in March 1996 was truly in full quiescence, 
and demonstrates that the source has returned to the same state after the 
outburst which began in 
April 1996 (Orosz et al. 1997).  
The $JK$ light curves are similar in shape to 
the optical curves.  
The increased scatter in the $K$ band is consistent with the increased 
scatter in the comparison stars, and does not require intrinsic source
variability. 

We find a deviant high point corresponding
to HJD $\sim 2,451,476.5$.  In all cases but the $V$ band, data from
this date lie significantly above the rest of the light curve, possibly
due to  
a flare or a 
hot spot.  This is unexpected, given the small contribution of the
accretion flow.  We removed this point for purposes of model fitting.  

In order to find an improved value for the orbital period we combined our 
data with the OB97 quiescent data (Figure 2).  We were unable to phase
the combined data set on the OB97 spectroscopic period of 
$2.62157 \pm 0.00015$ days.  However, OB97 used both quiescent and 
outburst radial velocity data to 
calculate this period.  
As pointed out by \citet{ShvdH99}, X--ray heating of the
secondary star can cause the `effective center' of the secondary to shift,
making a sinusoidal fit inadequate.  The OB97 period is therefore
suspect. 
Using our data set in conjunction with only the photometric data set of 
OB97, 
we found a preferred photometric period of $2.62191 \pm 0.00020$ days 
from the Phase
Dispersion Minimization \citep{S78} package in IRAF. 
The Lomb-Scargle \citep{PR89} and Clean \citep{RLD87} algorithms within the 
Period package \footnote{http://www.starlink.rl.ac.uk/star/docs/sun167.htx/sun167.htmlstardoccontents}
gave similar results.  This period is 
consistent with the published photometric period of $2.62168 \pm 0.00014$
days of \citet{vdH98}.  

\section {Discussion} 

Our primary purpose in measuring multiwavelength quiescent light 
curves of SXT's is to 
constrain the disk contribution in quiescence.  This is not only interesting 
in itself but provides a crucial input into the determination of inclination 
and therefore to the determination of the mass of the primary as well as 
other binary system parameters.

Past work on GRO J1655--40 has consistently found a model with 
an extended cool disk to best fit the data.  
In particular, the depth of the minimum at 
phase 0.5 (OB97 convention) in the $B$ light curve could not be fit without 
a grazing eclipse of the star by the disk.  Since the size of such an eclipse 
is a very strong function of inclination, very tight limits on
$i$ were derived (OB97, van der Hooft et al. 1998).  
At the same time, the small 
relative contribution of light from the disk in the optical wavebands 
required a very cool outer disk.  
The current work has improved on previous studies in two significant ways.
We use an improved modeling code which uses model atmosphere 
tables to compute local intensities, providing significantly improved
limb darkening calculations.  In addition we have obtained infrared data, 
allowing us to better constrain the contribution of the outer parts 
of the disk, which are cooler than the star.  

We used the recently developed ELC code (Orosz \& Hauschildt
2000, hereafter OH2000) to model the light curves.  This
code, which is loosely based on the code described in OB97, uses model
atmosphere specific intensities rather than specific intensities computed
using black bodies and a one- or two-parameter limb darkening law. For the
present problem we used the {\sc NextGen} grid for cool giants \citep{Hau99}. 
The {\sc NextGen} models are the most
comprehensive and detailed models available for cool stars.  The model
atmospheres are computed using spherical geometry, rather than the usual
plane-parallel approximation.  OH2000 show that for low gravity stars
($\log g\lesssim 3.5$) the limb darkening behavior of the extended
spherical models deviates significantly from a simple linear law.  
Since the secondary star in GRO J1655-40 has a relatively low gravity
($\log g\lesssim 3.5$ in general, and formally reaching zero at the $L_1$ 
point itself)  the consequences of the nonlinear limb darkening are
likely to be important.  In general, the {\sc NextGen} models predict
a lower intensity near the limb of the star.  The overall result is that in
most cases the amplitude of the ellipsoidal light curve is {\em larger}
when the NextGen intensities are used compared to the light curve computed 
using black body intensities (see OH2000 for a detailed discussion of this
point).  This effect is maximized in the bluest band passes,
where a small downward shift in temperature makes the biggest decrease in
flux.  Note that 
it is precisely the $B$ band lightcurve near phase $0.5$ which OB97
suggest requires the partial eclipse.

Our primary conclusion based on modeling the new data with the new code
is that the large, cool disk and 
partial eclipse advocated
in our previous study (OB97) and that of \citet{vdH98} is no longer 
viable.  Indeed, we find no evidence for {\it any} contribution from the 
accretion flow to the optical/IR lightcurves.
In Figure 1 we have plotted our data with the large disk model (computed
with the improved model atmosphere code) which provides the best fit to 
the OB97 data.  
While this model fits our optical data as well as
it fits the OB97 data, it fails dramatically in the IR, where 
the large contribution from the cool disk  
dilutes the ellipsoidal light curves too much.
In fact, a model with an eclipsing cool outer disk would require an outer 
disk temperature on the order of $100~K$ as found by \citet{vdH98}.  We argue
below that such a low outer disk temperature is physically implausible, due
to the radiation field of the F--star companion.
Alternatively, all bandpasses are well fit by a model with no disk 
contribution or eclipses (see Fig. 2).
The two models (with or without disk) are statistically indistinguishable
in terms of their fit to 
the $B$ band at phase 0.5.  There is still a small residual discrepancy, but it
is similar in size for both models, in contrast with the results
of OB97 
who found that with their cruder modeling
of limb darkening the partial eclipse model fit significantly better.  

A large cool disk with an outer temperature of $100~K$ also seems 
improbable given the proximity 
of an F-star companion.  While it is beyond the scope of this paper to solve
the detailed equations of gas dynamics and radiative transfer to determine
the outer disk temperature, it seems reasonable to derive an approximate 
lower limit on this temperature due to the radiation field of the companion
star.  The ratio of flux from the companion star at the
point of the disk nearest the $L_1$ point to that at the surface of the 
star itself is approximately ${(R_{star}/D_{disk})}^2$ where $R_{star}$ is the 
effective radius of the secondary star calculated using the Eggleton 
approximation \citep{E83} and $D_{disk}$
is the closest distance between the center of the secondary star and the outer
edge of the accretion disk.  Thus the black-body temperature generated by the
star's radiation field when it impacts the outer edge of the (presumably 
optically thick) disk would be 
\begin{displaymath}
T_{edge} = \sqrt{\frac{R_{star}}{D_{disk}}} \times T_{star}.  
\end{displaymath}
where $T_{edge}$ is the temperature due to the radiation field of the star at 
that point and $T_{star}$ is the mean surface temperature of the star.  Since 
all the gas in the outermost regions of the disk originated on the
secondary star and continues to be irradiated by the 
secondary at least some of the time as it orbits the primary, $T_{edge}$ 
may provide a rough lower limit to $T_{outer}$.
We emphasize that the actual value of of $T_{outer}$ depends on the 
temperature of the gas as it leaves the L1 point (which may be considerably
cooler than the average temperature of the star), the gravitational 
potential energy gained by the gas as it falls toward the disk, and 
any density change or thermal energy transfer which may occur upon the 
impact of the gas with the accretion disk.  

Figure 3 displays limits on the extent and outer temperature 
of the quiescent disk.  The allowed region is bounded at low $T_{outer}$ 
by $T_{edge}$.  
An upper limit on $T_{outer}$ can be derived empirically by computing the
limits on the disk contamination of the ellipsoidal lightcurves.  We have
done this by 
approximating the accretion flow as a
black--body disk with a single temperature equal to that at the outer edge.
This provides an appropriate limiting case, as any temperature rise toward
the center of the disk will increase the total flux contribution of the 
disk resulting in more stringent limits.  
We model the disk flux contribution as a constant dilution factor to the 
ellipsoidal lightcurves, taking into account the system inclination and 
an assumed outer disk radius.  The f-statistic is used to compare the
resulting lightcurves with our data, and determine whether they
are statistically different at the one and three $\sigma $ level.
Table 1 lists the allowed 
one and three $\sigma$ disk dilution limits for each filter computed 
from the model fits.
The curves shown in Figure 3 are derived from the $V$ band data, which provided
the tightest constraint on the allowed disk region.
Note that the limits for the combined light curves 
are much more severe than for any one filter, both because there are 
many more data points involved and because any contribution from the disk 
would be maximized in one particular bandpass, and thus would have to be 
much lower in most of the others.
Finally, Figure 3 also shows the disk size at which partial eclipses of
the star by the disk would occur.
Note that this is a somewhat porous limit, since it varies with $i$ and 
$Q$, but it demonstrates the qualitative effect of the lack of evidence
for such eclipses.

Based on the arguments presented above, we adopted a model which consists
only of ellipsoidal variations from the Roche-lobe filling secondary.  
Hence the only free parameters are the mass ratio $Q$ and the inclination
$i$.  We adopted a spectral type of F6III \citep{ShvdH99}, which
corresponds to a mean secondary star temperature of $T_{\rm eff} = 6336$~K 
\citep{Gr92}.  We had three basic binary system observables, namely the
photometric light curves in the five filters, the radial velocity curve
\citep{ShvdH99}, and the measurement of the mean projected
rotational velocity of the secondary star v$_{\rm rot}\sin i$ \citep{Isr99}.  
All of these observables were obtained when the source was
in complete X-ray quiescence ($L_x \leq 10^{-3}L_{\rm opt}$).  Thus the
light curves and the velocity curve did not suffer any biases caused by
X-ray heating.  It is well known that tidal distortion of the
secondary star can cause distortions in the line profiles from which the
rotational velocity is measured, leading to a systematic error in
v$_{\rm rot}\sin i$ (e.g.\ Kopal 1959; Marsh, Robinson, \& Wood 1994;
Shahbaz 1998; OH2000). However, the measurement of the rotational velocity
was obtained at the photometric phase 0 (i.e. the closest approach of the
F-star), which is the phase when the potential distortions to the line
profiles are minimized.  We computed numerical broadening kernels for
phase 0 and compared them to the analytic kernel commonly used and found
almost no systematic biases.  We conclude that the \citet{Isr99} value 
of of v$_{\rm rot}\sin i$ represents the true mean projected rotational
velocity of the F-star.

Our fitting procedure makes use of all of the binary system observables
and their uncertainties.  We will analyze three different data sets together.
First are the photometric light curves  reported in this
paper.  There are a total of 227 observations in the five filters
(48, 48, 47, 43 and 41 in B, V, I, J, and K respectively).  The second
data set is the radial velocity curve of Shahbaz et al.\ (1999), which
has 29 points.  Finally, the single measurement of 
$v_{\rm rot}\sin i=93\pm 3$ kms$^{-1}$ provided by Israelian et al.\ (1999)
is our third data set.  We seek a binary system model which best fits all three basic
observables simultaneously, i.e.\ we minimize the total $\chi ^2$:
$\chi ^2_{\rm tot}=\chi ^2 _{\rm photo}+\chi ^2 _{rm RV} + \chi ^2 _{\rm rot}$.

There is  some ambiguity in assigning weights to the various observations when 
fitting several quantities simultaneously.  In this case, we simply assigned
equal weights to each of the three data sets.  For the data sets to be weighted
equally requires that the values of $\chi ^2$ associated with the best fit to
each one be the same.
A three-parameter sinusoid fit
to the Shahbaz et al.\ (1999) radial velocities yields $\chi ^2_{\rm RV}/\nu = 0.98$.
We therefore take the quoted uncertainties on each radial velocity measurement at
face value.  We also adopted the $1 \sigma $ uncertainty in $v\sin i$ quoted by
Israelian et al.\ (1999).  In the case of our photometry, however, using the observational
uncertainties derived from comparison stars resulted in $\chi ^2/\nu > 1$ for our
best fits.  Therefore we have arbitrarily increased the errors used for each point
by a factor of $1.27$, which yields $\chi ^2/\nu =1$ for the best fits.

We then defined a grid of points in the $Q$-$i$ plane where 
$2.0 \le Q \le 3.65$ in steps of 0.05 and $66^{\circ} \le i \le
74.8^{\circ}$ in steps of $0.2^{\circ}$.  At each point in this grid the
orbital separation $a$ was adjusted to minimize the total $\chi^2$.
The orbital separation is needed to convert the model velocity curve and 
rotational velocity into physical units, and is also needed for the
computation of the light curve, since the model atmosphere 
intensities are tabulated as a function of the gravity in physical
units.

Using this procedure, we find a 
minimum overall $\chi^2_{\rm tot}$ at $Q=2.6$ and
$i=70.2^{\circ}$ (see Table 2).  
This point and $\chi^2$ contours denoting the one 
and three $\sigma$ confidence limits are shown in Figure 4.
Combining the allowed regions in $Q$ and $i$ shown in Figure 4
with the \citet{ShvdH99} mass function  $f(M)$, 
we find the probability distribution 
for $M_1$ shown in Figure 5.
To obtain this curve we divided the $Q-i$ plane into 50 confidence 
regions corresponding to the 
probability associated with $\chi^2_{\rm tot}$.  Then we selected equal 
numbers of $Q, i$ pairs
in each confidence region, and for each $Q, i$ pair chose values of $f(M)$
based on the \citet{ShvdH99} value and error.
A more detailed discussion of this procedure may be found in \citet{OW99}. 
The probability density shown in Figure 5 is obtained by calculating an 
$M_1$ for each point in the
$(Q,i,f(M))$ space, and summing the resulting values
of $M_1$ weighted by an appropriate likelihood.  The likelihood $L_{Qif}$ 
of each
combination of $Q$, $i$, and $f(M)$ is determined by
$$
L_{Qif}=\exp [-\chi ^2_{Qi} - \chi ^2_f]
$$
where $\chi ^2_{Qi}$ is the reduced $\chi ^2$ determined for the relevant
values of $Q$ and $i$ as described above, and $\chi _f = (f-f_m)/\sigma _f$,
where $f_m$ is the value of $f(M)$ measured by \citet{ShvdH99} and
$\sigma _f$ is their quoted error.  The distribution shown in Figure 5 is
then the normalized probability distribution created by summing all the
values of $M_1$ weighted by their likelihoods.   
We find a preferred $M_1$ value of 
$6.3 \pm 0.5\,M_{\sun }$ ($95\%$ confidence).  The peak of this 
distribution is lower than the result quoted in OB97 due to 
the lower value of $f(M)$ determined by \citet{ShvdH99}, and 
the distribution is wider than the errors quoted by OB97 
because the limits on the inclination are less stringent without
the requirement for a partial eclipse.

\section{Summary}

We present $BVIJK$ quiescent light curves of the SXT GRO J1655--40.  
The shape of our light curves is consistent with that of previously 
published results, supporting the assumption that the source truly went
into quiescence in 1996 before its most recent outburst.
We report a refined photometric period of $2.62191 \pm 0.00020$ days.  
We find that our best fit to the 
data at all wavelengths is provided by a model with no contribution from an
accretion flow. For this fit we find a range in {\it i} and Q 
shown in Figure 4, which is consistent with, but somewhat
wider than,  previous findings.
We also find a probability distribution of $M_1$ shown in Figure 5 where
 $M_1=6.3 \pm 0.5\,M_{\sun }(95\%$ confidence).
We have placed limits on the temperature and physical size of any accretion
disk component.

\acknowledgments

This research was supported by NSF grant AST--9730774.  Construction of the 
ANDICAM was carried out by the Ohio State University astronomical 
instrumentation group and sponsored by the NSF grant AST--9530619. 
We are grateful for
conversations with Darren Depoy and Rick Pogge about the performance of the 
instrument, and with Nick Suntzeff about the calibration of the IR channel.  
David Gonzalez and Juan Espinoza provided their usual expert services as 
mountain-top obvservers, and Suzanne Tourtellotte managed the data processing
with her usual aplomb.  \citet{ShvdH99} were kind enough to provide the 
RV data.   We thank Elene Terry for development
of the data processing software and Raj Jain for 
his software expertise and endless patience.

\clearpage

\figcaption[fig1.ps]{Our $BVIJK$ light curves folded on $P_{orb}=2.62191$ days.
The open circles represent the high point which was removed from the data for 
analysis purposes.  Optical magnitudes were found to be consistent 
with those of OB97.  Overplotted are the best fit OH2000 models to
the OB97 data set.  This model required a partial eclipse by a disk
with $T_{outer}\approx 1000$K.
Note the poor fit in $J$ and $K$ due to the excess disk contribution in
the infrared.
\label{fig1}}

\figcaption[fig2.ps]{Our light curves (filled circles) are shown with the 
quiescent light curves of OB97 (filled triangles).  Overplotted (lines)
is the best fit model with no disk contribution.  Below
is shown a comparison stars of similar brightness (filled squares) 
to the program object and the residuals to the fit (filled circles).  The 
scatter in the residuals matches that in the comparison stars, except 
near phase 0.5 in the $B$ where the limb darkening effects are at a
maximum.}

\figcaption[fig3.ps]{Upper and lower limits to the size and temperature of 
a thin disk accretion flow.  The thin lower line is $T_{edge}$
which we argue represents a lower
limit to the 
temperature of the outer disk due to illumination by the secondary star.  
The dashed vertical line 
represents the limit imposed by the lack of grazing eclipses due to 
an infinitely thin disk, at $i\,=\,70.2 \pm 1.9\arcdeg$ and $Q\,=2.6 \pm 0.3$.
This should be seen as an approximate limit due to its
sensitivity to the chosen mass ratio and specific shape information. 
The upper curves represent the $1 \sigma$ (bold line) and $3 \sigma$ 
(dashed line) upper limits on spatial extent and temperature of a
single temperature black--body disk which is assumed to contribute a 
constant flux dilution to the ellipsoidal light curves.  The limits are 
derived from the $V$ band data, which provided the most stringent limits on 
the allowed disk size and temperature.}

\figcaption[fig4.ps]{Allowed region in $Q-i$ space. 
Contours shown are the one and three $\sigma$
 limits imposed both 
by fits to the photometric data and by the constraint imposed by the
measured v$_{\rm rot} \sin i$ of \citet{Isr99}, as described in
the text.  The region above the dotted line would result in X--ray eclipses of
the central source, which are not observed.}

\figcaption[fig5.ps]{The probability distribution of $M_1$, found by 
weighting the values $M_1$ derived
over a range of $Q$, $i$, and $f(M)$ by an appropriate likelihood,
as described in the text.}

\clearpage

\begin{deluxetable}{lll}  
\tablecolumns{3}  
\tablewidth{0pc}  
\tablecaption{Allowed Disk Dilution}  
\tablehead{  
\colhead{Filter} & \colhead{$1 \sigma$}   & \colhead{$3 \sigma$}} 
\startdata  
$B$ & $21\%$  & $28.5\%$\\
$V$ & $11.5\%$ & $17.5\%$\\
$I$ & $19.5\%$ & $25\%$ \\
$J$ & $15\%$ & $24.5\%$\\
$K$ & $30\%$  & $46\%$\\
 
\enddata  
\end{deluxetable}

\clearpage

\begin{deluxetable}{lll}  
\tablecolumns{3}  
\tablewidth{0pc}  
\tablecaption{Orbital Parameters of GRO J1655-40}  
\tablehead{  
\colhead{Parameter} & \colhead{Result}   & \colhead{Reference}} 
\startdata  
Orbital Period, photometric (days) & $2.62191\pm0.0002$ & This paper\\
$K_2$ (km $s^{-1}$) & $215.5\pm2.4$ & \citet{ShvdH99}\\
v$_{\rm rot}$sin$i$ & $93.0 \pm3.0$ & \citet{Isr99} \\
Mass Function $f(M)/M_{\sun}$ & $2.73\pm0.09$ & \citet{ShvdH99} \\
Q & $2.6 \pm 0.3\,(95\%$ confidence) & This paper -- see Fig. 4 \\
Inclination $i$ (deg) & $70.2 \pm 1.9\,(95\%$ confidence) & This paper -- see Fig. 4 \\ 
Primary Mass $M_1/M_{\sun}$ & $6.3 \pm 0.5\,(95\%$ confidence) & This paper -- see Fig. 5 \\
Secondary Mass $M_2/M_{\sun}$ & $2.4\pm 0.4\,(95\%$ confidence) & This paper\\
Secondary Radius $R_2/R_{\sun}$ & $5.0\pm 0.3\,(95\%$ confidence) & This paper\\
\enddata  

\end{deluxetable}  

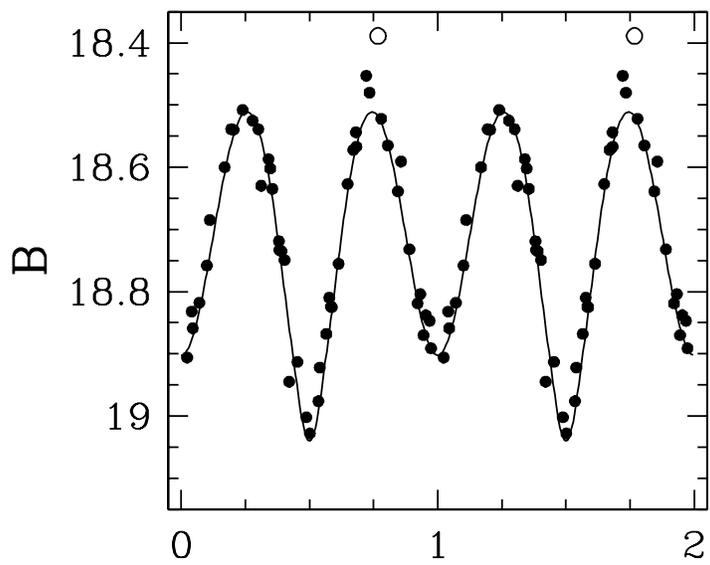
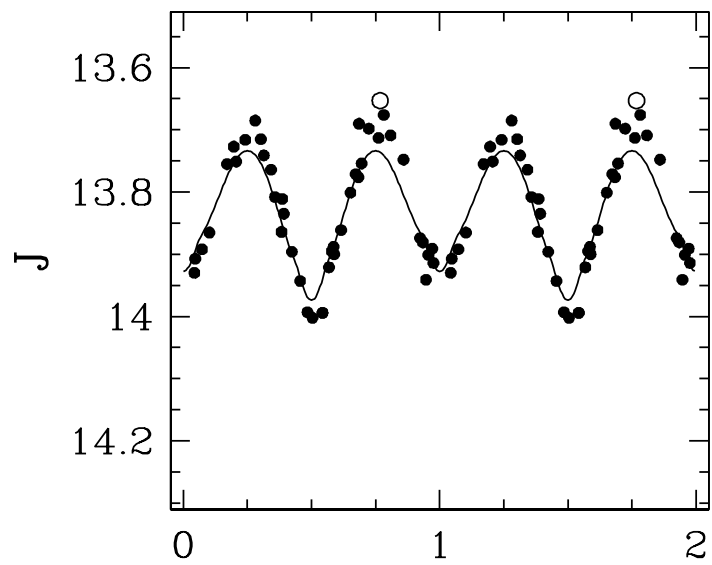
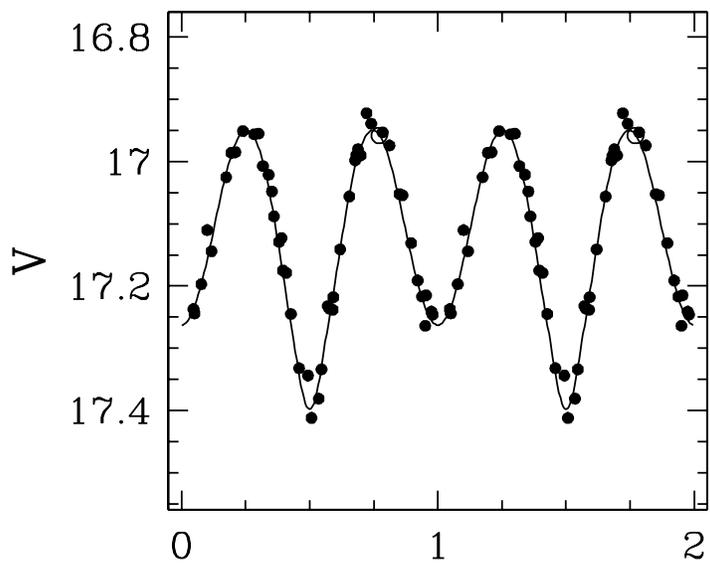
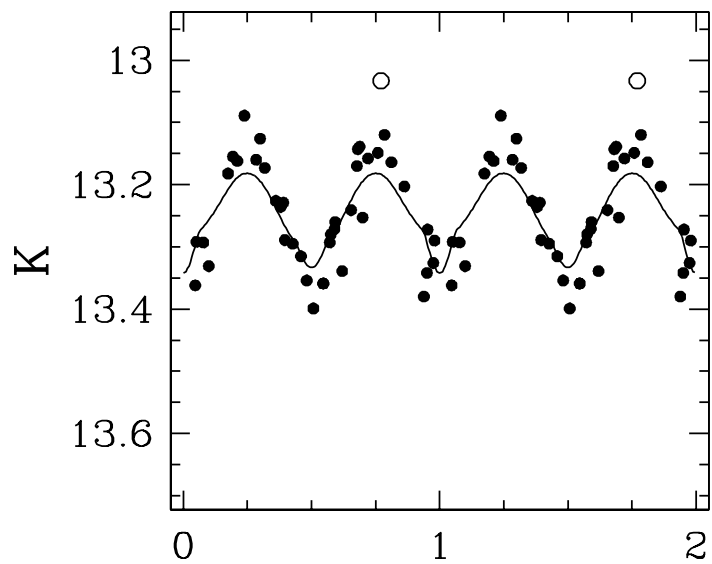
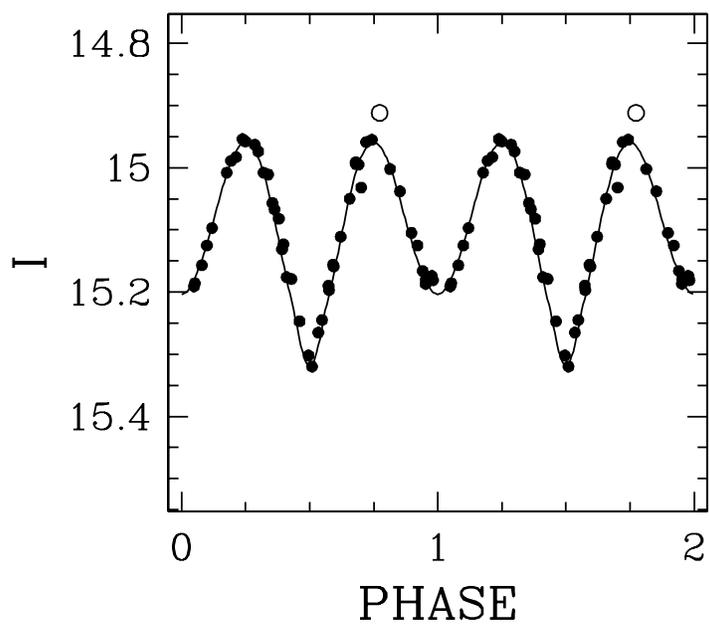
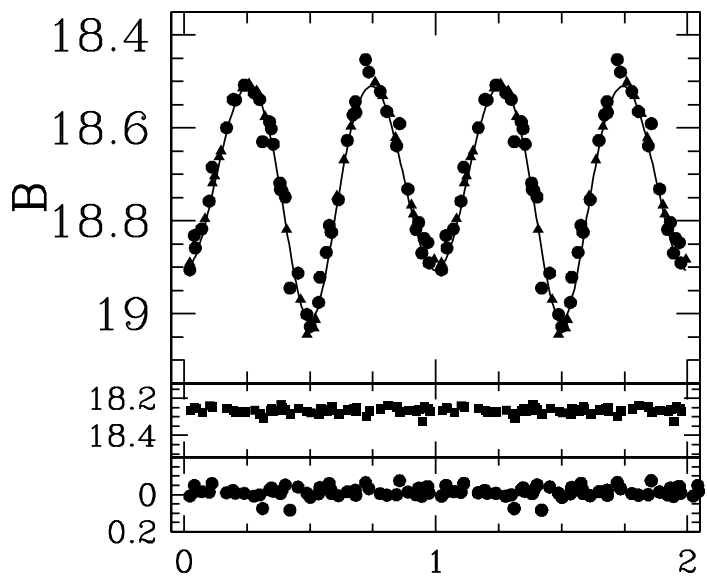
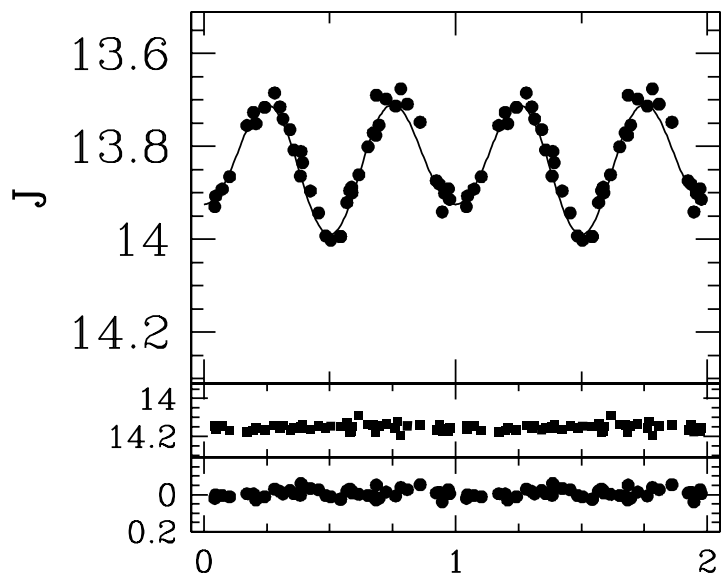
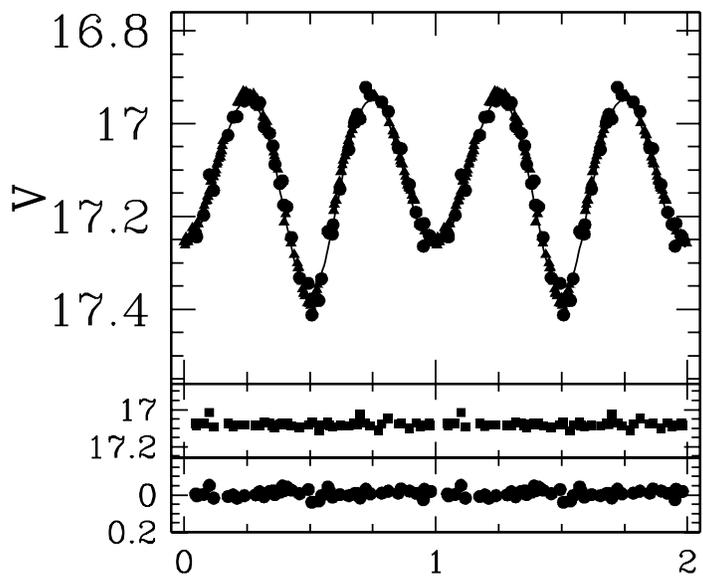
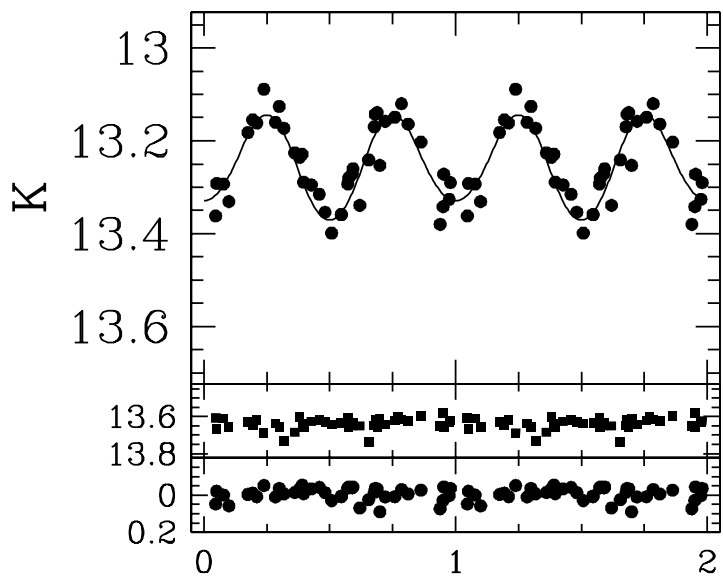
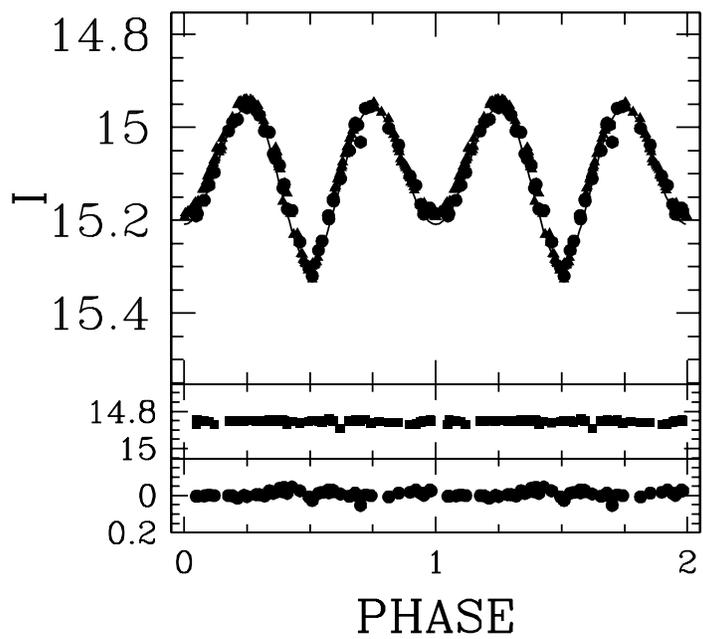
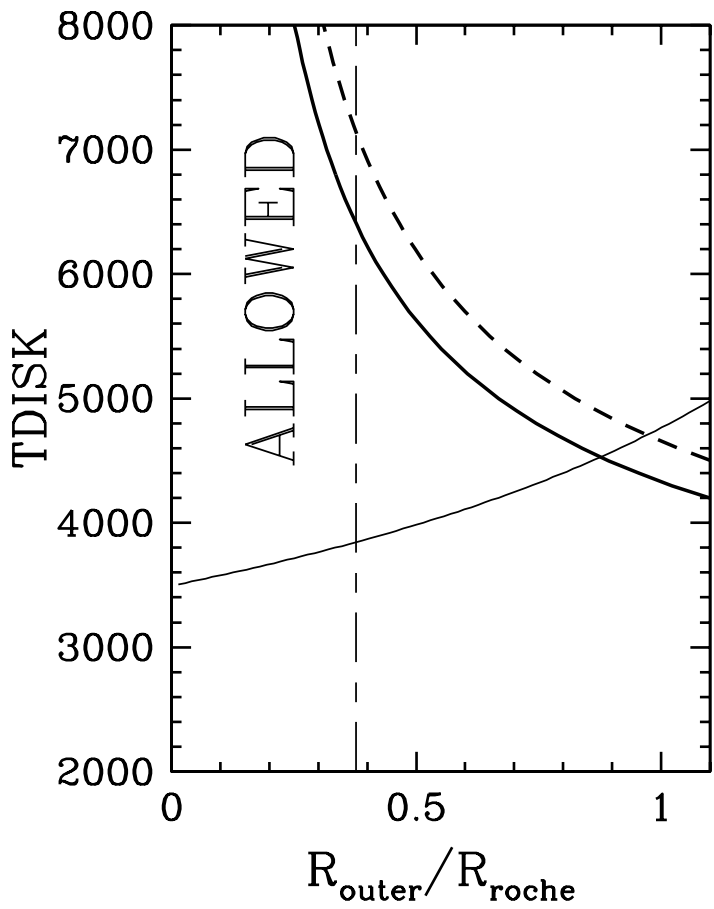
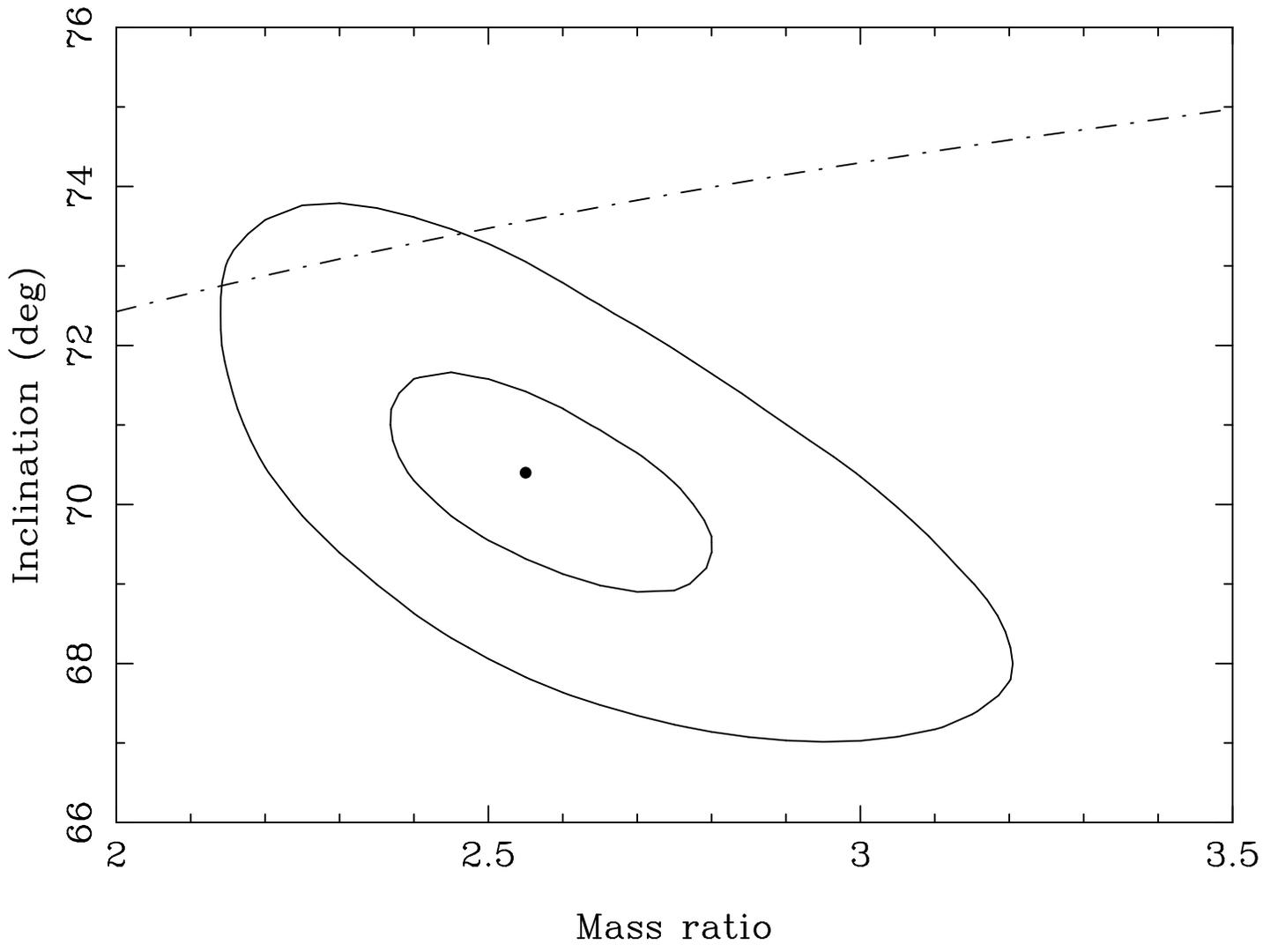
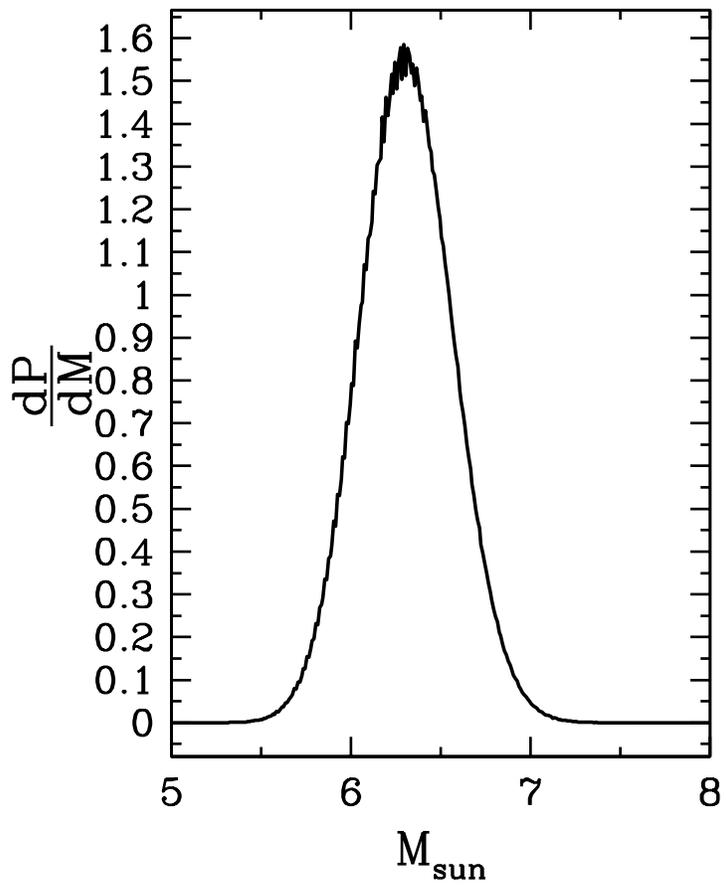
\end{document}